\documentstyle[12pt]{article}

\def\noi{\noindent}

\def\del{\partial}

\def\dis{\displaystyle}

\begin{document}

\begin{center}
{\Large\bf Third quantization of $f(R)$-type
gravity II \\ - General $f(R)$ case -}
\\[10mm]
\end{center}

\noi
\hspace*{1cm}\begin{minipage}{14.5cm}
Yoshiaki Ohkuwa$^1$ and Yasuo Ezawa$^2$\\

\noi
$^1$ Section of Mathematical Science, Dept. of Social Medicine, Faculty 
of Medicine,\\
$\ $  University of Miyazaki, Kihara 5200, Kiyotake-cho, 
 Miyazaki, 889-1692, Japan\\
$^2$ Dept. of Physics, Ehime university, 2-5 Bunkyo-cho, 
Matsuyama, 790-8577, \\ $\ $ Japan\\

\noi
Email : ohkuwa@med.miyazaki-u.ac.jp, ezawa@phys.sci.ehime-u.ac.jp 
\\

%\noi
%Short title :  Third quantization of $f(R)$-type gravity II
%\\

\noi
{\bf Abstract}\\[2mm]
In the previous paper we 
examined the third quantization of the $f(R)$-type gravity   
and studied the Heisenberg uncertainty relation of the universe 
in the example of $f(R)=R^2$. 
In this work the  Heisenberg uncertainty relation 
of the universe is investigated in the general $f(R)$-type gravity 
where tachyonic states are avoided. 
It is shown that, at late times 
namely the scale factor of the universe is large, the spacetime 
becomes classical, and, at early times namely the scale 
factor of the universe is small, the quantum effects dominate.
\\

\noi
\\

\end{minipage}

\section{Introduction}

Investigations of $f(R)$-type gravity arises from two main motivations.
In early stage, such theories were interested in because of their theoretical 
advantages. The theory of graviton is renormalizable\cite{UDW}\cite{Stelle}. 
It seems to be possible to avoid the initial singularity of the universe 
predicted by the theorem proved by Hawking\cite{Hawking}\cite{NT}.
And inflationary model without inflaton field is possible\cite{Staro}.
Another motivation began 
after the discovery of the accelerated expansion of the universe\cite{accel} 
\cite{SN}\cite{WMAP} and large value of the dark energy density\cite{WMAP}, 
and such theories have been attracting 
much attention\cite{CDTT}\cite{SF}\cite{NO}\cite{BCNO}.

Quantum mechanical aspects of the theory are mainly applied to cosmology, 
namely quantum cosmology\cite{SS}, and black holes\cite{Faizal1}.
The fundamental equation describing the dynamics of the universe is the 
Wheeler-DeWitt equation(WDW eq.) which is the differential equation for 
the wave function of the universe\cite{WDW}.   
However, it is well known that, in general, 
WDW eq. has the problem 
that the probabilistic interpretation is difficult as in the case of 
the Klein-Gordon equation.  
One of the proposed ideas to solve this problem is the third quantization 
in analogy with the quantum field 
theory \cite{Banks}\cite{Rubakov}\cite{McGuigan}\cite{Giddings}\cite{Hosoya}
\cite{Fischler}\cite{Xiang}\cite{Pohle}\cite{Abe}\cite{Horiguchi}
\cite{Castagnino}\cite{Vilenkin}\cite{Ohkuwa}\cite{Pimentel}
\cite{Faizal1}\cite{Faizal2}\cite{Calcagni}\cite{Faizal3}. 
Then the third-quantized universe theory describes a system of many universes.
Third quantization is useful to describe bifurcating universes and merging
universes,  
if an interacting term is introduced in the Lagrangian for the 
third quantization.

The quantum cosmology of the $f(R)$-type gravity using WDW eq. 
 has already been studied \cite{SS}. 
As noted above the third quantized version is desirable, 
and the third quantization of $f(R)$-type gravity was also investigated 
in Ref.\cite{Faizal1}.   
However, in it black holes were studied but cosmology was not treated.  
So in the previous work we examined the third quantization of the 
$f(R)$-type gravity, 
using explicit form of the action which yields WDW eq.  
of $f(R)$-type gravity, 
and we investigated the Heisenberg uncertainty relation of the universe 
in the example of $f(R)=R^2$ \cite{3QfR}. 
In this work we investigate the uncertainty relation of the universe 
in the general $f(R)$-type gravity where tachyonic states are avoided.

We start from the effective theory of the $f(R)$-type gravity in a flat 
Friedmann-Lemaitre-Robertson-Walker metric. 
Then a suitable change of variable is performed and WDW eq. is written down. 
Quantizing this model once more, we obtain the third-quantized theory 
of $f(R)$-type gravity. 
The Heisenberg uncertainty relation is investigated 
in the general $f(R)$-type gravity where tachyonic states are avoided.  
It will be shown that, at late times  
namely the scale factor of the universe is large, the spacetime 
becomes classical, and, at early times namely the scale 
factor of the universe is small, the quantum effects dominate.

In section 2, the effective theory of $f(R)$-type gravity in the case of
 a flat Friedmann-Lemaitre-Robertson-Walker metric is summarized .
In section 3, the third quantization of this theory is considered.
In section 4, the uncertainty relation is studied 
in the general $f(R)$-type gravity 
where tachyonic states are avoided.
Summary is given in section 5.

\section{Generalized gravity of $f(R)$-type}

Generalized gravity of $f(R)$-type is one of the higher curvature 
gravity in which the action is given by
$$
S=\int d^4 x{\cal L}=\int d^4 x\sqrt{-g}f(R).                      \eqno(2.1)
$$
The spacetime is taken to be $4$-dimensional. 
Here $g\equiv \det g_{\mu\nu}$ and $R$ 
is the scalar curvature.

Let us consider the next action
$$
S=\int d^4 x \sqrt{-g}\Bigl[f(\phi)+f'(\phi)(R-\phi)\Bigr]  ,      \eqno(2.2)
$$
where $f'(\phi)=\dis{d f(\phi) \over d \phi}$ and we assume 
$f''(\phi) \neq 0$. 
It is well known that the field equations of this action are 
obtained from the field equations for (2.1)  
in which $f(R)$ is replaced with $f(\phi)$ and the following equation 
$$
R=\phi .                                                           \eqno(2.3)
$$
If we substitute Eq.(2.3) into Eq.(2.2), we formally obtain Eq.(2.1) 
\cite{SF}.

In order to make things simple, let us consider the case of a flat 
Friedmann-Lemaitre-Robertson-Walker metric \cite{SFf(R)}, 
$$
ds^2=-dt^2+ a^2 (t) \sum_{k=1}^3 (dx^k)^2          .               \eqno(2.4)
$$
Then the scalar curvature is written as 
$$
R= 6 \left({\ddot{a} \over a} + {\dot{a}^2 \over a^2} \right) ,    \eqno(2.5)
$$
with $\dot{a}=\dis{da \over dt}$.

Using Eqs.(2.4),(2.5), we can straightforwardly transform Eq.(2.2) to 
$$
S_{eff}=\int d^4 x \Bigl[ a^3 f(\phi)-6f''(\phi)\dot{\phi}a^2 \dot{a}
-6f'(\phi) a {\dot a}^2 -f'(\phi)\phi a^3 \Bigr]
=\int d^4 x {\cal L}_{\it eff}                                ,    \eqno(2.6)
$$
where the partial integration has been applied to the term containing 
$\ddot{a}$ \cite{SFf(R)}\cite{SS}.

Then standard canonical formalism leads to the Hamiltonian written as \cite{3QfR} 
$$
{\cal H}_{\it eff} =\dis -{p_a p_{\phi} \over 6a^2 f''(\phi)}
+{f'(\phi)p_{\phi}^2 \over   6a^3f''(\phi)^2}
+a^3 f'(\phi) \phi - a^3 f(\phi)          .                        \eqno(2.7)    
$$
The time reparametrization invariance means 
$$
{\cal H}_{\it eff} = 0 .                                           \eqno(2.8)
$$                                                                
This equation and the field equations of motion from (2.6) 
give those of (2.1) and Eq.(2.3) in the case of Eq.(2.4). 
Therefore ${\cal L}_{\it eff}$ can be regarded as the effective Lagrangian 
for Eq.(2.1) when the metric is given by Eq.(2.4) 
\cite{SFf(R)}\cite{SS}\cite{3QfR}.

\section{Third quantization}

Eqs.(2.7),(2.8) lead to WDW eq.\cite{SS} 
whose kinematic terms are rather complicated, 
and it is difficult to obtain the action for the third quantization. 
Therefore, in this section we first make a change of a variable 
to make the kinematical terms simpler, 
and then we derive WDW eq.
Next we write down the action for the third quantization which yields 
WDW eq. as the field equation.
Then we carry out the third quantization.

Now let us make the change of a variable as follows:
$$
\varphi=\varphi(\phi) \equiv \ln f'(\phi), \qquad f'(\phi)=e^{\varphi}, 
\qquad 
\phi=f'^{-1}(e^{\varphi})   .                                      \eqno(3.1)
$$
Then the Lagrangian is expressed as 
$$
{\cal L}_{\it eff}=-6a {\dot a}^2e^{\varphi}
-6a^2 {\dot a}e^{\varphi}{\dot \varphi}
-a^3f'^{-1}(e^{\varphi}) e^{\varphi}
 +a^3 f\Bigl( f'^{-1}(e^{\varphi}) \Bigr)  ,                       \eqno(3.2)
$$
and the Hamiltonian constraint becomes \cite{3QfR}
$$
{p_{\varphi}^2 \over a}-p_a p_{\varphi}
+6 a^5 f'^{-1}(e^{\varphi}) e^{2 \varphi}
-6a^5 e^{\varphi}  f\Bigl( f'^{-1}(e^{\varphi}) \Bigr) 
=0                                                      .          \eqno(3.3)
$$
Substituting
$$
p_a \rightarrow -i {\del \over \del a} ,\qquad
p_{\varphi} \rightarrow -i {\del \over \del \varphi} ,             \eqno(3.4)
$$
we obtain WDW eq. 
$$
-{1 \over a}{\del^2 \psi \over \del \varphi^2}
+{\del^2 \psi \over \del a \del \varphi}
+6a^5 f'^{-1}(e^{\varphi}) e^{2 \varphi} \psi
-6a^5 e^{\varphi} f\Bigl( f'^{-1}(e^{\varphi}) \Bigr) \psi=0.      \eqno(3.5)
$$
Here $\psi(a, \varphi)$ is the wave function of the universe.

Now let us comment on the possibility of tachyonic states in this WDW eq. 
In order to examine Eq.(3.5) in the Klein-Gordon form, 
we make change of variables as 
$$
\tau = a +\varphi +\ln a ,\quad \sigma = a -\varphi -\ln a  .      \eqno(3.6)
$$ 
Then we obtain 
$$
{\del^2 \psi \over \del \tau^2}-{\del^2 \psi \over \del \sigma^2} 
+ U \psi = 0 ,                                                     \eqno(3.7)
$$
where
$$
\begin{array}{ll}
U&=\dis 6a^5 f'^{-1}(e^{\varphi}) e^{2 \varphi}
-6a^5 e^{\varphi}  f\Bigl( f'^{-1}(e^{\varphi}) \Bigr), \\[3mm] 
&=6\left( {\tau+\sigma \over 2} \right)^3 
f'^{-1}\biggl({2 \over \tau+\sigma} e^{\tau-\sigma \over 2}\biggr) 
e^{\tau-\sigma}
-6\left( {\tau+\sigma \over 2} \right)^4
f\left( f'^{-1} \biggl( {2 \over \tau+\sigma} 
e^{\tau-\sigma \over 2}\biggr) \right) e^{\tau-\sigma \over 2} . 
\end{array}                                                        \eqno(3.8)
$$
From Eqs.(3.6) we notice that $\tau$ can be considered as 
the time variable, since $\tau$ is a monotonic increasing function 
of the scale factor $a$ . 
Therefore, in order to avoid tachyonic states, 
$U \geq 0$ is required, since $U$ is the square of the effective 
mass \cite{Abe}. 
The condition $U \geq 0$ means
$$
f'(R) \Bigl( f'(R) R -f(R) \Bigr) \geq 0 ,                         \eqno(3.9)
$$
in the original variables.
Notice that this condition is satisfied, for example,
$f(R)=R^2$ , when $R \geq 0$ , 
and $f(R)={1 \over 16\pi G}R+ c R^2$ , where $c$ is a 
small positive constant .

The action for the third quantization 
to yield WDW eq.(3.5) can be written as 
$$
\begin{array}{ll}
S_{3Q}&=\dis \int d a d \varphi {1 \over 2} \left[
{1 \over a} \Biggl( {\del \psi \over \del \varphi} \Biggr)^2
-{\del \psi \over \del a}{\del \psi \over \del \varphi}
+6a^5 f'^{-1}(e^{\varphi}) e^{2 \varphi} \psi^2 
-6a^5 e^{\varphi} f\Bigl( f'^{-1}(e^{\varphi}) \Bigr) \psi^2
\right]   , \\[3mm]
&=\int d a d \varphi {\cal L}_{3Q}  . 
\end{array}                                                        \eqno(3.10)
$$ 
If we consider $a$ to be the time coordinate from now on, 
the canonical momentum which is conjugate to $\psi$ is written as
$$
p_{\psi}={\del {\cal L}_{3Q} \over \del 
(\del \psi / \del a)}
=-{1 \over 2}{\del \psi \over \del \varphi} .                      \eqno(3.11)
$$ 
The Hamiltonian is given by
$$
\begin{array}{ll}
{\cal H}_{3Q}&=\dis{ p_{\psi}{\del \psi \over \del a}-{\cal L}_{3Q}} , \\[5mm]
&=\dis{-{2 \over a}p_{\psi}^2-3a^5 f'^{-1}(e^{\varphi}) e^{2 \varphi} \psi^2
+3a^5 e^{\varphi} f\Bigl( f'^{-1}(e^{\varphi}) \Bigr) \psi^2} .
\end{array}                                                        \eqno(3.12)                       
$$

In order to third quantize this system we impose the equal time 
commutation relation 
$$
\bigl[ {\hat \psi}(a,\varphi) , {\hat p_{\psi}}(a,\varphi') \bigr] 
= i \delta (\varphi - \varphi')  .                                 \eqno(3.13)
$$
We use the $\rm Schr\ddot{o}dinger$ picture, so we take the operator 
${\hat \psi} (a,\varphi)$ as the time independent c-number field 
$\psi (\varphi)$, and we substitute the momentum operator as 
$$
{\hat p_{\psi}} (a,\varphi) \rightarrow 
-i{\del \over \del \psi (\varphi)} .                               \eqno(3.14)
$$
Then we obtain the $\rm Schr\ddot{o}dinger$ equation
$$
i{\del \Psi \over \del a} ={\hat {\cal H}}_{3Q} \Psi   ,           \eqno(3.15)
$$
that is
$$
i{\del \Psi \over \del a} 
=\left[ {2 \over a} \Biggl( {\del \over \del \psi (\varphi)} \Biggr)^2
-3a^5 f'^{-1}(e^{\varphi}) e^{2 \varphi} \psi^2 (\varphi) 
+3a^5 e^{\varphi} f\Bigl( f'^{-1}(e^{\varphi}) \Bigr) \psi^2 (\varphi) 
\right] \Psi ,                                                     \eqno(3.16)
$$
where $\Psi$ is the third quantized wave function of universes.

\section{Uncertainty relation}

In order to estimate the uncertainty, 
we assume the Gaussian ansatz for the solution to the 
$\rm Schr\ddot{o}dinger$ equation (3.16) as is often done 
$$
\Psi(a,\varphi,\psi (\varphi) )
=C \exp \left\{ -{1 \over 2}A(a,\varphi)[\psi (\varphi)-\eta (a,\varphi)]^2
+iB(a,\varphi)[\psi (\varphi)-\eta (a,\varphi)] \right\}  ,        \eqno(4.1)
$$
where $A(a,\varphi)=D(a,\varphi)+iI(a,\varphi)$ 
\cite{Floreanini}\cite{Pohle}\cite{Horiguchi}\cite{Pimentel}\cite{3QfR}.    
The real functions 
$D(a,\varphi), I(a,\varphi),$  $B(a,\varphi)$ and $\eta (a,\varphi)$ 
should be determined from Eq.(3.16).
$C$ is the normalization of the wave function.
The inner product of two functions $\Psi_1$ and $\Psi_2$ is defined as 
$$
\langle \Psi_1 , \Psi_2 \rangle 
=\int d \psi (\varphi) \Psi_1^*(a,\varphi,\psi (\varphi))
 \Psi_2(a,\varphi,\psi (\varphi))          .                       \eqno(4.2)
$$

The Heisenberg uncertainty relation can be calculated as \cite{3QfR}
$$
(\Delta \psi (\varphi))^2 (\Delta p_{\psi}(\varphi))^2
={1 \over 4} \Biggl( 1+ {I^2 (a,\varphi) \over D^2 (a,\varphi)} 
\Biggr)   ,                                                        \eqno(4.3)
$$
where
$(\Delta \psi (\varphi))^2 \equiv \langle \psi^2 (\varphi) \rangle
-\langle \psi (\varphi) \rangle^2$ 
and
$(\Delta p_{\psi}(\varphi))^2 \equiv \langle p_{\psi}^2 (\varphi) \rangle
-\langle p_{\psi}(\varphi) \rangle^2$ .

Note that to evaluate (4.3), only $A(a,\varphi)$ 
is necessary. 
Substituting the ansatz (4.1) into 
the $\rm Schr\ddot{o}dinger$ equation(3.16), we obtain
$$
-{i \over 2} {\del A(a,\varphi) \over \del a}
={2 \over a} A^2 (a,\varphi) 
+3a^5e^{\varphi} \left[ f \Bigl(f'^{-1}(e^{\varphi}) \Bigr)
-e^{\varphi} f'^{-1}(e^{\varphi}) \right]  .                       \eqno(4.4)
$$
Writing
$$
\ln a = {\alpha \over 6}  ,                                        \eqno(4.5)
$$
we have
$$
-3i{\del A(\alpha,\varphi) \over \del \alpha} 
=2 A^2 (\alpha,\varphi) 
+3 e^{\alpha} e^{\varphi} \left[ f \Bigl(f'^{-1}(e^{\varphi}) \Bigr)
-e^{\varphi} f'^{-1}(e^{\varphi}) \right] .                        \eqno(4.6)
$$
In order to solve this equation, let us write
$$
A(\alpha,\varphi)={3i \over 2}{\del \ln u(\alpha,\varphi) 
\over \del \alpha} ,                                               \eqno(4.7)
$$
where $u(\alpha,\varphi)$ is a suitable function.
Then $u(\alpha,\varphi)$ must satisfy the equation, 
$$
{\del^2 u(\alpha,\varphi) \over \del \alpha^2}
+ k(\varphi) e^{\alpha} u(\alpha,\varphi) = 0 ,                    \eqno(4.8)
$$
where 
$$
k(\varphi)={2 \over 3}e^{\varphi}\Bigl[ e^{\varphi} f'^{-1}(e^{\varphi})
-f\Bigl( f'^{-1}(e^{\varphi}) \Bigr) \Bigr]  .                     \eqno(4.9)
$$
Now let us assume the condition (3.9) to avoid tachyonic states, 
and let us introduce a new variable 
$$
z=2\sqrt{k(\varphi) e^{\alpha} }     ,                             \eqno(4.10)
$$
which plays a role of time coordinate. 
Then we have 
$$
{\del^2 u(z,\varphi) \over \del z^2}
+{1 \over z}{\del u(z,\varphi) \over \del z} 
+ u(z,\varphi) = 0 .                                               \eqno(4.11)
$$
As this equation can be regarded as the ordinary 
differential equation with respect to $z$ 
%under the assumption that $\varphi$ is fixed other than in $z$,
with a parameter $\varphi$,   
this is the case when $\nu=0$ in the following Bessel's equation
$$
{d^2 u(z) \over d z^2}
+{1 \over z}{d u(z) \over d z} 
+\left( 1- {\nu^2 \over z^2} \right) u(z) = 0 .                    \eqno(4.12)
$$ 
Therefore we have the solution
$$
u(z,\varphi)=c_J(\varphi) J_0 (z) + c_Y(\varphi) Y_0 (z)      ,    \eqno(4.13)                        
$$
where $J_0, \ Y_0$ are the Bessel functions of order $0$ 
and $c_J, \ c_Y$ are arbitrary complex functions of $\varphi$. 

From Eqs.(2.3),(3.1),(4.5),(4.7),(4.9),(4.10),(4.13), we can obtain 
$$
z=2 \sqrt{{2 \over 3} f'(R)[f'(R)R-f(R)] }\ a^3 ,\qquad 
\varphi=\ln (f'(R))                                                \eqno(4.14)
$$ 
and 
$$
A(z,\varphi)=-i {3z \over 4} 
{c_J(\varphi) J_1 (z) + c_Y(\varphi) Y_1 (z) 
\over c_J(\varphi) J_0 (z) + c_Y(\varphi) Y_0 (z) },    
                                                                   \eqno(4.15)
$$
where we have used $J'_0(z)=-J_1(z), \ Y'_0(z)=-Y_1(z)$ \cite{AS}.

Since $A(z,\varphi)=D(z,\varphi)+iI(z,\varphi)$ , we have
$$
D(z,\varphi)= -{3i \over 4\pi 
\vert c_J(\varphi) J_0 (z) + c_Y(\varphi) Y_0 (z) \vert^2}
[c_J(\varphi) c_Y^*(\varphi) -c_J^*(\varphi) c_Y(\varphi)]  ,      \eqno(4.16)
$$
where we used 
$J_0(z)Y_1(z)-J_1(z)Y_0(z)=-\dis{2 \over \pi z}$ \cite{AS}, and 
$$
\begin{array}{ll}
I(z,\varphi)=&-\dis{3z \over 8  
\vert c_J(\varphi) J_0 (z) + c_Y(\varphi) Y_0 (z) \vert^2}\times \\[6mm]
&\Bigl[ 2\vert c_J(\varphi) \vert^2 J_0(z) J_1(z)
+2\vert c_Y(\varphi) \vert^2 Y_0(z)Y_1(z)  \\[3mm]
&+(c_J(\varphi) c_Y^*(\varphi) + c_J^*(\varphi) c_Y(\varphi))
(J_0(z) Y_1(z) + J_1(z) Y_0(z)) \Bigr]  .                          
\end{array}                                                       \eqno(4.17)
$$

So if we assume 
$c_J(\varphi) c_Y^*(\varphi) -c_J^*(\varphi) c_Y(\varphi) \neq 0$ 
(Note that in this case both of $c_J(\varphi), c_Y(\varphi)$ are nonzero.), 
we obtain 
$$
\begin{array}{ll}
\dis {I^2 (z,\varphi) \over D^2 (z,\varphi)}= 
&-\dis{\pi^2 z^2 \over 
4[c_J(\varphi) c_Y^*(\varphi) -c_J^*(\varphi) c_Y(\varphi)]^2}\times \\[6mm]
&\Big[ 2\vert c_J(\varphi) \vert^2 J_0(z) J_1(z)
+2\vert c_Y(\varphi) \vert^2 Y_0(z)Y_1(z)  \\[3mm]
&+(c_J(\varphi) c_Y^*(\varphi) + c_J^*(\varphi) c_Y(\varphi))
(J_0(z) Y_1(z) + J_1(z) Y_0(z)) \Bigr]^2  .                          
\end{array}                                                       \eqno(4.18)
$$

At late times namely $a \rightarrow \infty$
i.e. $z \rightarrow \infty$,
$$
\begin{array}{ll}
 J_0(z) &\sim \dis{ \sqrt{{2 \over \pi z}} \cos (z-{\pi \over 4})} ,
\qquad J_1(z) \sim \dis{ \sqrt{{2 \over \pi z}} \sin (z-{\pi \over 4})} , \\[5mm]
Y_0(z) &\sim \dis{\sqrt{{2 \over \pi z}} \sin (z-{\pi \over 4})} ,
\qquad Y_1(z) \sim \dis{-\sqrt{{2 \over \pi z}} \cos (z-{\pi \over 4})} ,
\end{array}                                                     
$$
\cite{AS}
we have
$$
\begin{array}{ll}
\dis{I^2 (z,\varphi) \over D^2 (z,\varphi)}  &\sim 
-\dis{\left[( \vert c_J(\varphi) \vert^2-\vert c_Y(\varphi) \vert^2) \cos (2z)
+(c_J(\varphi) c_Y^*(\varphi) +c_J^*(\varphi) c_Y(\varphi)) \sin (2z) \right]^2 
\over [c_J(\varphi) c_Y^*(\varphi) -c_J^*(\varphi) c_Y(\varphi)]^2} \\[3mm]
&\sim O(1) . 
\end{array}                                                      \eqno(4.19)
$$
This and Eq.(4.3) mean that at late times   
namely $a \rightarrow \infty$, 
it is plausible that the spacetime becomes classical 
in the sense that the quantum fluctuations become minimum.  

On the other hand at early times namely
$a \rightarrow 0$ i.e. $z \rightarrow 0$,
$$
\begin{array}{ll}
\dis J_0(z) &\sim \dis{1-{z^2 \over 4}} ,
\qquad J_1(z) \sim \dis{{z \over 2}} ,  \\[5mm]
Y_0(z) &\sim \dis{{2 \over \pi} \ln z}  , \ 
\qquad Y_1(z) \sim \dis{-{2 \over \pi z}} , 
\end{array}                          
$$
\cite{AS}
we obtain
$$
{I^2 (z,\varphi) \over D^2 (z,\varphi)} \sim 
-{16 \vert c_Y(\varphi) \vert^4 \over 
\pi^2 [c_J(\varphi) c_Y^*(\varphi) -c_J^*(\varphi) c_Y(\varphi)]^2}
( \ln z )^2 \sim \infty  .                                       \eqno(4.20)
$$
This and Eq.(4.3) mean that the fluctuation of the third quantized universe 
field becomes large at early times namely 
$a \rightarrow 0$. 
Therefore the quantum effects dominate for the small values of the 
scale factor of the universe.

\section{Summary}

In this work the third quantization of the $f(R)$-type gravity is investigated, 
when the metric is a flat Friedmann-Lemaitre-Robertson-Walker one.
The Heisenberg uncertainty relation of the universe 
is investigated in the general $f(R)$-type gravity 
where tachyonic states are avoided.  
It has been shown  that, at late times 
namely the scale factor of the universe is large, the spacetime 
becomes classical, and, at early times namely the scale 
factor of the universe is small, the quantum effects dominate.
This result is similar to Ref.\cite{Abe}\cite{Horiguchi}\cite{Pimentel}\cite{3QfR} 
but is not similar to Ref.\cite{Pohle}, 
where it was shown that quantum effects dominate 
also when the scale factor is large.  
However, as pointed out in Ref.\cite{Horiguchi} this era corresponds to the 
classically forbidden region in that model\cite{Pohle}.

As a future work, 
though our formulation started from the effective action (2.2), 
it will be interesting to quantize (2.1) directly as in 
Ref.\cite{Faizal1}, since the quantization of the $f(R)$ -type gravity 
would not be unique.

\section*{Acknowledgements}

One of the authors (Y.O.) would like to thank Prof. H. Ohtsuka 
for letting know him a clue to solve Eq. (4.8).

\end{document}